\documentclass[12pt,a4paper]{article}
\usepackage{amssymb}
\usepackage{amsmath}
\usepackage{cite}
\usepackage{graphics}
\usepackage[dvips]{color}
\usepackage[dvips]{graphicx}
\usepackage{float}
\usepackage[caption = false]{subfig}

\newcommand{\beq}{\begin{equation}}
\newcommand{\bea}{\begin{eqnarray}}
\newcommand{\bec}{\begin{center}}
\newcommand{\eeq}{\end{equation}}
\newcommand{\eea}{\end{eqnarray}}
\newcommand{\eec}{\end{center}}
\newcommand{\vev}[1]{\ensuremath{\langle #1 \rangle}}

\newcommand{\so}{\ensuremath{SO(10) \;}}

\newcommand{\wur}{\dfrac{1}{2 \, \pi}}
\def\ga{\mathrel{\raise.3ex\hbox{$>$\kern-.75em\lower1ex\hbox{$\sim$}}}}
\def\la{\mathrel{\raise.3ex\hbox{$<$\kern-.75em\lower1ex\hbox{$\sim$}}}}
\newcommand{\mgut}{\ensuremath{M_{GUT}\,}}
\newcommand{\meff}{\ensuremath{M_{eff}\,}}
\newcommand{\meffth}{\ensuremath{M_{eff}(th) \,}}
\newcommand{\meffexp}{\ensuremath{M_{eff}(exp)\,}}

\newcommand{\astrong}{\ensuremath{\alpha_{strong}\,}}

\begin{document}

\begin{titlepage}  
\begin{flushright} 
\parbox{4.6cm}{UA-NPPS/BSM-10/02\\
              }
\end{flushright} 
\vspace*{8mm} 
\begin{center} 
{\large{\textbf {
High Energy Thresholds in SUSY GUTs - A refined analysis}}}\\
\vspace{14mm} 
{\bf A. ~\ Katsikatsou}  

\vspace*{6mm} 
  {\it University of Athens, Physics Department,  
Nuclear and Particle Physics Section,\\  
GR--15771  Athens, Greece}

\end{center} 
\vspace*{15mm} 

\begin{abstract}

In a previous paper we have developed a novel method in order to parametrize the effect of the large number of HETs into as fewer parameters as possible. Apart from its obvious advantages this parametrization serves as a vehicle for the examination of the validity of a minimal \so model concerning a series of constraints. Among them is demand for unification of the gauge couplings, the experimental values of the strong coupling constant \astrong and lower experimental bound of the proton lifetime. All of these claims lead to preferred regions both in the soft the superheavy parameter space. In this paper we give the necessary updates of our results which stem mainly from the recent experimental measurements. We also include some additions in our analysis involving the universal trilinear coupling $A_0$ and the rate $x,$ which is related to superheavy vevs. Finally we cross-check the preferred regions of the parameter space and narrow the even more by applying new constraints based on the results of the LHC experiment for the mass of the Higgs particle and the supersymmetry exclusion limits.

\end{abstract} 
\vspace{2.5cm}

\hspace{1cm}\hrule width 7.5 cm 

\vspace{.2cm}

\hspace{.2cm} {E-mail\,:\, kkatsik@phys.uoa.gr}

\end{titlepage} 
 \clearpage
\newpage

\baselineskip=17pt

\section{Introduction}\label{model}



The supersymmetric grand unified theories (SUSY GUTs) is the most natural extension of the Standard Model. Their most important feature is the unification of the three gauge couplings in a very large energy scale which is called unification scale \mgut. These theories provide some very interesting and fundamental predictions, among them the nucleon decay and the quantization of the electric charge. Despite these good features, the GUTs possess some disadvantages as well, with the main one being the big number of superheavy parameters which are necessary to define the high energy thresholds (HETs) \cite{Weinberg,Hall}. Having as a goal the easiest and more effective control of these parameters, we introduce a new method \cite{Katsikatsou:2010hj}, according to which the effect of the HET is comprised within fewer free parameters produced from the initial group of the superheavy parameters.

In order to check the efficiency of the new method that we created, we apply it to an \so\,SUSY GUT model \cite{Georgi}. In the context of this model the $2/3$ splitting is realized through the Dimopoulos-Wilczek mechanism. The SUSY \so\,GUT models reveal many good characteristics, such as the inclusion of all the fermions of the same generation into only on non-irreducible representation of \so $\mathbf{16}$ and the inherent see-saw mechanism which gives the expected tiny masses to the Standard Model neutrinos \cite{neutrinos}. The later is related to the mechanism of baryogenesis which in \so\, is based on thermal leptogenesis \cite{Fukugita:1986hr}. Moreover the predictions for the proton lifetime are still in accord with the experimental limits.
In SUSY \so\,GUT models the idea of Yukawa couplings unification arises as a natural consequence \cite{Yukawauni} and the $R$ parity is an intrinsic symmetry of the theory, leading to many positive results with the most important being the stability of the LSP particle.  

This paper is organized as follows.
In sec.\ref{so10model}, we give a brief description of the adopted \so model.
In sec.\ref{sechet}, we recall the proposed method for the collective treatment of the HETs and their impact on the running of the gauge coupling from \mgut to $M_Z.$ We explain the practicalities of its implementation. We specify the boundary conditions needed, in the context of CMSSM and we set the fundamental constraint of gauge coupling unification.
In sec.\ref{secproton}, we discuss the other important and quite strict constraint of our analysis, the nucleon instability, through the lower experimental bound of the proton lifetime. The basis of this constraint is a mass parameter \meff, which depends on the HETs of the GUT model and the gauge coupling constants at the electroweak scale.
The results of our analysis are presented in sec.\ref{results1}. We test the model and the method on its ability to produce gauge coupling unification and also against the experimental limitations from electroweak precision measurements and proton lifetime, as well as the recent results from the LHC experiment, concerning the Higgs and the supersymmetric particles masses.
Our conclusions are given in ~{sec.\ref{conclusions}.}

\section{The \so model} 
\label{so10model}

We apply our method to a minimal supersymmetric \so  model first introduced in \cite{Barr:1}, whose low energy effective theory is the constrained MSSM. 
The Higgs content of the particular model consists of one Higgs multiplet $A$ in the adjoint $\mathbf{45}_H$ representation and two pairs of $\mathbf{16}_H \, + \, \overline{\mathbf{16}}_H$ multiplets, named $C \, + \, \overline{C}, \; C' \, + \, \overline{C'}$, in the spinorial representation. It also contains two Higgs fields $T_1, \, T_2$ in the vector $\mathbf{10}_H$ representation and several \so singlets. Because of these content an additional $U(1)\times Z_2 \times Z_2$ symmetry is needed.
%
%
%
The Higgs superpotential responsible the \so breaking \cite{Barr:1, Barr:2} is of the form
$$W \, = \, W_A \, + \, W_C \, + \, W_{ACC'} \, + \, W_{2/3}  \quad .$$
The adjoint sector, $W_A$, is responsible for the Dimopoulos - Wilczek vev of $A$ \cite{DWmechanism}:  
\beq
\vev{A} \, = \, diag(\alpha, \,\alpha, \,\alpha, \, 0, \, 0) \, \otimes \, i \tau_2, 
\label{vevA}
\eeq
which breaks $SU(5)$ to the Standard Model symmetry preserving $U(1)_{B - L}.$
The parameter $\alpha$ is of the order of the GUT breaking scale $M_{GUT}.$ The spinor sector, $W_C$, of the Higgs superpotential forces the pair of spinor Higgs multiplets $C \, + \, \overline{C}$ to get superheavy vevs along the $SU(5)$ singlet direction. In this way, the rank of the group is reduced from 5 to 4, since the $B - L$ symmetry is broken. The adjoint-spinor coupling terms, $W_{ACC'}$, are necessary to prevent the manifestation of colored pseudo-Goldstone bosons with small masses. The presence of the later fields may destroy the unification of gauge coupling constants and the low energy particle spectrum. Finally, the $W_{2/3}$ term is responsible for doublet-triplet splitting. For further details on the terms of the superpotential refer to \cite{Katsikatsou:2010hj}.

%
%


\section{Gauge Couplings Unification and High Energy \\Thresholds} \label{sechet}

The strictest constraint to which we submit the previously described model is the unification of all three gauge couplings to a very large energy scale called \mgut. Therefore, while solving numerically the 2-loop RGEs for the gauge coupling constants $\alpha_i \, (i = 1, 2, 3)$ from this scale \mgut to the electroweak scale and vice versa, we
impose the gauge coupling unification condition at \mgut:  
\beq
\alpha_1 (\mgut) \, = \, \alpha_2 (\mgut) \, = \, \alpha_3 (\mgut) \, \equiv \, \alpha_G \, ,
\label{unif}
\eeq
At \mgut, we also activate the universal boundary conditions for the soft supersymmetry breaking parameters induced by gravity mediated SUSY-breaking: 
$$
m_i \, = \, m_0 \quad , \quad  M_i \, = \, M_{1/2} \quad , \quad A_i \, = \, A_0 \,
$$
and hence we operate within the CMSSM.
Our analysis is not further restricted by imposing Yukawa unification as well.

However, we trigger another, novel boundary condition which regards the 1-loop HETs effect of the superheavy mass spectrum. From the evolution of $\alpha_i$ from \mgut to the lowest high energy threshold,  $M_L$, 
one can derive, ignoring for the moment the two loop effects, that 
\beq
\alpha_i^{-1} (M_L) \, = \, \alpha_G^{-1} (\mgut) \, + \, 
\wur \, b^{GUT} \, \ln \dfrac{\mgut}{M_L} \, + \,
\wur \, \sum_{H_k} \, b_i^{H_k} \, \ln \dfrac{M_L}{m_{H_k}} \,  ,
\label{runmxml}
\eeq
where $b^{GUT}$ is the sum of all the beta function coefficients of SM, SUSY and superheavy particles and $b^{H_k}$ are the beta function coefficients of every superheavy particle. 
We define
\beq
\alpha_G^{-1} (M_L) \, \equiv  \, \alpha_G^{-1} (\mgut) \, + \, 
\wur \, b^{GUT} \, \ln \dfrac{\mgut}{M_L} \, , 
\label{agutml}
\eeq
which represents the running from \mgut to $M_L$, if HETs are ignored, and 
\beq
c_i \, \equiv \, 
\wur \, \sum_{H_k} \, b_i^{H_k} \, \ln \dfrac{M_L}{m_{H_k}} 
\label{cidef}
\eeq
Then the equation (\ref{runmxml}) emerges in the form:
\beq
\alpha_i^{-1} (M_L) \, = \, \alpha_G^{-1} (M_L) \, + \, c_i \, .
\label{MLcon}
\eeq
This is the new boundary condition for the three gauge coupling constants at the lowest HET, $M_L$ that takes into account the effects of all HETs, which are included within the $\, c_i\,$s.

For any set of the model's parameters, $p_j$, we assign a \textquotedblleft vector\textquotedblright \, in a five dimensional space 
$\, \vec{c} = ( c_1, \, c_2, \, c_3,  \,  \, M_L,  \, M_{GUT} \, ) $, which includes, besides the $c_i$s, the values of the maximum, \mgut, and lowest high energy mass, $M_L \,$. 
Further, for the purpose of utilizing $c_i$s as inputs, we use a random sample generator, which assigns random numbers to each GUT parameter $p_j$. In this way, random points $\vec{p} \equiv (p_1, p_2, ...p_N)  $ are drawn in the model parameter space and each of this is mapped to a $\, \vec{c} \,$ defined before. 

The other couplings of the theory and the soft supersymmetric mass parameters are also influenced by the HETs and the boundary conditions mentioned.
In order to treat their HETs in a collective way, using quantities similar to $c_i \,$, we run for each coupling or mass parameter $F$ the two-loop RGEs without the contribution of the superheavy particles and we include the effect of the 1-loop HETs, at the end by duly correcting each derived quantity $F$ at the scale $M_L$ as 
$$
F^{cor}(M_L) \,=\, F (M_L) \,+\, \Delta_F \,,
$$
where
$$
\Delta_F \,=\,  \dfrac{\alpha_G^2}{4\, \pi} \,  \ln \left( \dfrac{M_{GUT} }{ M_L} \right) \,\sum_i \,G_i(M_{GUT}) \,
\left[ \ln \, \left( \dfrac{M_{GUT} }{ M_L} \right) \, 
b_i^H \, + \, 2 \, \pi \, c_i \right] \, .
$$
This approach is valid provided that the lowest, $M_L$, and the highest, $M_{GUT}$, threshold are not separated by more than three orders of magnitude: 
\beq 
10^{-3} \, < \,\frac{M_L}{\mgut} \, < \, 1 \label{constr_ml}
\eeq
From the random samples, we find that  
on an average $\log \frac{M_L}{\mgut} \, \simeq \, - \, 2.7$ and thus the approximation is more than satisfactory.

In our analysis, we also take into account the low energy thresholds from the supersymmetric fields and heavy SM particle. We embed them in the evolution of the 2-loop RGEs through boundary conditions at $M_Z,$ as it is described in \cite{Katsikatsou:2010hj,Lahanas, Pierce:1996zz}.

The advantages of the new parametrization (\ref{cidef}) are:
\begin{enumerate}
\item
Instead of dealing with a large number of GUT parameters and masses, we only have a few to consider in our analysis, namely 
$\, c_1, \, c_2, \, c_3,  \,  \, M_L,  \, M_{GUT} \, $,
which define  $\, \vec{c} \,$ 
and carry all information on the the masses of the non-singlet superheavy fields of the model which contribute to the HETs in the running from $M_Z$ to $M_L$ and vice versa.
\item
The random  procedure  actually makes a selection by mapping the parameter space to a rather confined  region, at least in the \so model, which is 
spanned by the vectors $\, \vec{c} \,$. Then, within this region, points satisfying the experimental criteria  can be sought and the region shrinks even more.
Consequently, one avoids time-consuming scans over a multidimensional (10-dimensional in \so ) parameter space, since the random  procedure  has already selected the points $\, \vec{c} \,$ which meet the criteria. 
\item
The RG equations are solved without the inclusion of HETs and their effect is properly taken into account by changing accordingly the boundary conditions at $M_L$ and at the unification scale. 
\item
Our analysis is also constrained by the relation (\ref{constr_ml}) which makes the used scheme fast and accurate.
\item
This method is applicable to almost every GUT model and is particularly useful in the cases of numerous and complex superheavy mass spectrums.
\end{enumerate}
 
In this way we have found a very convenient way to parametrize the effect of HET, which is applicable to almost any GUT model, using the variables $\, \vec{c} \,$.

\section{The Proton Decay constraint}
\label{secproton}

For every SUSY GUT model, the nucleon instability is a primary prediction (for recent but not exhaustive list of studies \cite{Mohapatra1,Nath-Costa,resentso10,Li:2010dp} and for a broad overview see \cite{Nath2}) and it is caused mainly by $D = 5$ operators \cite{Dimopoulos, Ellis, Nath, Yanagida}. The dominant decay mode is $p \, \rightarrow \, \overline{\nu} \, K^+$. The the most stringent limits on the proton lifetime are provided by the Super-Kamiokande experiment \cite{Regis:2012sn}. Currently the lower experimental bound is \cite{Hewett:2012ns}
\beq
 \tau(p \rightarrow \bar \nu K^+  ) > 4 \times 10^{33} \; {\rm yrs}.
\label{plifetime}
 \eeq
This is quite a restrictive value which rules out simple unification models such as the simple SUSY $SU(5)$ GUT model. 
For the \so model which we follow, the decay rate results from:
\beq
\Gamma( p \, \rightarrow \, \bar\nu \, K^+ ) \; = \; \sum_{i \, = \, e, \, \mu, \, \tau} \Gamma( p \, \rightarrow \, \bar\nu_i \, K^+ ) \, ,
\label{pdwidth} 
\eeq
with όπου  $i \, = \,e, \, \mu, \, \tau$.
Each of the partial rates in (\ref{pdwidth}) are given by \cite{Nath1, Nath, Yanagida}:
\beq
\Gamma( p \, \rightarrow \, \bar\nu_i \, K^+ ) \; = \; \biggl( \dfrac{\beta_p}{M_{eff}} \biggr)^2  \; |A|^2 \; |B_i|^2 \; C \, .
\label{width}
\eeq
The factor $\beta_p$ denotes the hadronic matrix element between the proton and the vacuum state of the three quark operator \cite{bp1}. Its most reliable calculation is given by using lattice QCD combined with chiral Lagrangian techniques \cite{Aoki:2006ib, Aoki:2008ku}.

Continuing on the rest of the factors in (\ref{width}): the $Α$ contains the quark masses at 1 GeV and \cite{Ellis} the CKM matrix elements \cite{pdb}, as well as $A_S$ and $A_L$ which represent the short-range renormalization effects between GUT and SUSY breaking scales \cite{Ellis, Yanagida} and the long-range renormalization effects between SUSY scale and 1 GeV \cite{Ellis} respectively.
We use the values $A_L \, = \, 1,4$ and $A_S \, = \, 2$ \cite{Nath2}, with which \cite{Hisano:2000dg, EmmanuelCosta:2003pu} also agree.
The $B_i$s are functions which describe the supersymmetric dressing of the proton decay diagrams and finally $C$ \cite{C,Nath1} contains chiral Lagrangian factors, which convert a Lagrangian involving quark fields to the effective chiral Lagrangian involving mesons and baryons \cite{chiral2}. 
For further details and values used in (\ref{width}) see \cite{Katsikatsou:2010hj}.

The proton decay rate is inverse proportional to an effective, non-physical, mass parameter, stemming from the Yukawa terms of the superpotential as a combination of superheavy Higgs particle:
\beq
M_{eff} \, = \, \frac{M_3 \, M_3'}{M_2} \, = \, \frac{(\lambda \alpha)^2}{M_2} \, ,
\label{meff}
\eeq
where $M_3 \, M_3'$ are the superheavy color triplet Higgs masses of the vector Higgs multiplets $T_1, \, T_2$. 
Thus, the expected proton lifetime turns out to be proportional to $M_{eff}^2$. 

At the same time, this effective mass is proved to be a function of the model's high energy thresholds and also the three gauge coupling constants at $M_Z.$ This correlation constitutes the basis for another severe constraint which our model has to face: the proton decay constraint originated from the experimental lower bound to the proton lifetime. This constraint is determined by the following relation:
\beq
\meffth > \meffexp \, \equiv \, \beta_p \, | A | \, \sqrt{  \tau_p \,  C \, \sum_i \, {| B_i | }^2      }.
\label{prtconstr}
\eeq
We call \meffexp the mass parameter which we extract from eq.(\ref{width}). 
On the right hand side of (\ref{prtconstr}) $\tau_p$ is the minimum of ({plifetime}) and the rest of the factors are explained previously in
({width}). We derive \meffth from:
\beq
\frac{M_{eff}}{M_Z} \, = \,\, \mathrm{e}^{h ( {\alpha'}_i^{-1} )} \,
f(x).
\label{yan_ours}
\eeq
The information of the all the HETs is incorporated within $f$, which in this particular \so model depends on only one dimensionless parameter $x$ defined as:
\beq
x \, \equiv \, \dfrac{\alpha}{ 2 \, c} \, ,
\label{sh_param}
\eeq
with $\alpha$ and $ c$ being the vevs of the superheavy Higgs fields in the adjoint and in the spinorial representation respectively.
In \cite{Katsikatsou:2010hj} we confirmed that a large $x$ satisfies easier the proton decay constraint. However, $x$ is naturally expected to be of $\mathcal{O}(1) \,$  as being the ratio of vevs which are both of order  $\sim \mgut$. On these grounds, $x$ cannot be taken  arbitrarily large.
The effect of the gauge coupling constants at $M_Z$ forms:
\beq
h ( {\alpha'}_i^{-1} ) \, =  \dfrac{5 \, \pi}{6} \, \bigl[ \, 3 \, {\alpha'}_2^{-1} (M_Z) \, - \, {\alpha'}_1^{-1} (M_Z) \, - \, 2 \, {\alpha'}_3^{-1} (M_Z) \, \bigr]. 
\eeq
In this function, the $\alpha'_i$s arise from the evaluation of the 1-loop RGEs for the gauge coupling constants ${\alpha}_i$, in the $\overline{DR}$ scheme, from the GUT scale $M_{GUT}$ down to the electroweak scale $M_Z,$ as we have already discuss in the previous section. For the running of the RGEs, we take into account the 1-loop high energy thresholds of the superheavy spectrum, as well as the low energy thresholds of all sparticles and heavy SM particles. The later is fulfilled by imposing low energy boundary conditions at $M_Z$. The $\alpha'_i$s also include a correction term which coordinates the numerically calculated from 2-loop RGEs gauge coupling constants of our analysis with the gauge coupling constants needed in (\ref{yan_ours}).

\section{Analysis - Results} \label{results1}

\subsection{Numerical procedure}

To achieve our goals, we created the appropriate Fortran code.
First we produce a large number of random vectors $\vec{c}_{in}  \, \equiv \,( c_i^{\, in}, \, M_L^{\, in}, \, M_{GUT}^{\, in} \,),$ which meet the theoretical requirements of \so. All these points have a common $M_{GUT}^{\, in},$ which is an input in our analysis. We primarily adopt $\, M_{GUT} = 2 \cdot 10^{16} \, GeV.$ We also use as inputs the components $\, c_1^{in}, \, \, c_2^{in} $ and we solve numerically the 2-loop RGEs from $M_Z$ and upwards by importing trial values for the gauge coupling constants at $M_Z.$ Then we determine the value of the lowest HET $M_L,$ from the boundary condition (\ref{MLcon}), and that of $\alpha_G(M_L).$ From eq.(\ref{agutml}) we can also decide the value of $\alpha_G(M_{GUT}).$ We derive the 2-loop RGEs relations from \cite{Martin:1993zk} and we adjust them to our notation which is mainly the one of \cite{Ellis:1989jf}.

The value of $M_L,$ which we extract in this way is unavoidably different from $M_L^{in}$ of the vector $\, \vec{c}_{in} $ and the same goes for the value of the $c_3$ parameter. The later is specified by
\beq
c_3 \;=\;    \alpha_3^{-1} (M_L) \, - \, \alpha_G^{-1} (M_L)  \, ,
\label{c3}
\eeq
so as to satisfy eq.(\ref{MLcon}) for the coupling constant $\alpha_3.$ The $\, c_1^{in}, \, \, c_2^{in}$ parameters undergo small corrections as well. In the following subsection we will discuss in detail these variations.

The running of the RGEs is resumed until \mgut, where the soft universal boundary conditions and the unification of the gauge couplings are applied, with the 1-loop HETs taking into account from $M_L$ up to \mgut, as we have already described. Then the running continues backwards, until the electroweak scale, where we enforce the boundary conditions for the low energy thresholds at the gauge coupling constants. The iteration of this procedure stops once we succeed convergence of the output values of the Higgs and Higgsinos mixing parameters $m_3^2$ and $\mu$ respectively. 

As inputs we use the values of the $\alpha_{em},$ the Fermi coupling constant $G_F,$ the physical mass of the $Z$ boson $Μ_Ζ,$ the physical masses of the top and tau quark, as well as the running mass $m_b(m_b)$ in $\overline{MS}$ \cite{pdb}. In addition to \mgut, we also fix the parameter $x.$
The soft breaking parameters $m_0, M_{1/2}, Α_0,$ the value of $\tan \beta$ and the sign of the Higgsino mixing parameter $\mu$ are also inputs of our analysis. For the later we choose $\mu \, > 0,$ which is consistent with the experimental results for $(g \, - \, 2)_\mu$ and $B \longrightarrow s \, + \, \gamma,$ especially if it is combined with universal gaugino masses at \mgut. The minimization conditions are solved with all the one-loop effective potential corrections and the dominant two-loop QCD and top Yukawa corrections taken into account. The value of the $| \mu |$ parameter is then determined by the minimization conditions and $\tan \beta \,$ is input. Therefore, apart from the particular treatment of HETs, the procedure is the standard one encountered in the constrained MSSM models.


\subsection{Convergence of results}

As we mention in the previous section \ref{sechet}, every time we run the 2-loop RGEs we use as input a set of four parameters $\vec{c}_{in} \,=\, \{ c_1^{in}, c_2^{in}, c_3^{in}, M_L^{in} \}$, generated from the random sample. In order to achieve unification at the given \mgut we correct their values and in this way we produce a different set of the same parameters 
$\vec{c}_{fin} \, = \, \{c_{1}^{fin}, c_{2}^{fin}, c_{3}^{fin}, M_{L}^{fin} \}$. Thus, at the end, we get a 
final point, $\, \vec{c}_{\, fin} \,$, which is a successful point if it belongs to the set of the randomly generated vectors $\, \vec{c}\,$ or unsuccessful and hence discarded if it lies outside the region spanned by $\, \vec{c}\,$ points. The biggest deviation is observed in the case of $c_{3}.$  Actually, this  is expected since the $c_3$ parameter is strongly correlated to \astrong and is forced to make the strong coupling be compatible with the unification scale that is determined by the couplings $\alpha_{1,2} \,$. This is implemented by the shift of eq.  (\ref{c3}), resulting to $c_3^{fin}$, which is always towards higher values comparing to input $c_3^{in} \,$. This is also the case for $c_{1,2}^{in} \,$ and $M_L^{in}\,$, which are also shifted towards higher values but to a much lesser extend.


%
%

Therefore, we add a new constraint in our analysis stemming from the demand of convergence between the input points $\vec{c}_{in}$ and the output points $\vec{c}_{fin},$ for the model to be successful.
We perform a "second" run using as inputs the elements of the previously produced output groups and at the same time we demand the satisfaction of all the constraints we have previously set. 
In this run, we first check whether these points are subset of the initial randomly generated points. Obviously, only a subset of those survives, if it does at all, and this comprises the set of successful points.  The numerical evaluation is performed by defining the "distance" $ \chi_i \, $ of the point $ \vec{c}_{fin} \,$ from any $ \vec{c}_{in}^{\,i} \,$ of the randomly generated points, according to: 
$$
\chi_i \equiv  \, \left|\frac{{c}^{fin}_1 - {c}^{\, in, \, i}_1}{{c}^{\,in, \, i}_1} \right| \,+\, 
\left|\frac{{c}^{fin}_2 - {c}^{\, in, \, i}_2}{{c}^{\,in, \, i}_2} \right| \,+ \,
\left|\frac{{c}^{fin}_3 - {c}^{\, in, \, i}_3}{{c}^{\,in, \, i}_3} \right| \,+ \,
\left|\frac{{M_L}^{fin} - {M_L}^{\, in, \, i}}{{M_L}^{\,in, \, i}} \right| \, .
$$
We judge how far $\, \vec{c}_{fin} \,$ stands from the set of the randomly generated points from the minimum of these $\,\chi = min \{ \chi_i's \} .$ If it turns out to be equal to zero it means that the final point coincides with one of the random points that were initially created. 

In our analysis and with one million random points, we have found that points with  $\chi \leq  0.1$ are acceptable by the model. 
This analysis certainly depends on the SUSY inputs.  In the $m_0$ - $ M_{1/2}$ plane, successful points are found  for values of $m_0$ and $ M_{1/2}$ reaching roughly $1000$ GeV. If one keeps the higher end of $ M_{1/2}$ values constant, the restrictions are satisfied altogether for values of $m_0\,$ up to approximately $1200$ GeV. If we loosen the restriction for $\chi$ to $\chi \leq  0.2$, keeping the other two constraints untouched, our successful region matches the one described in the following subsection. This findings supports not only our method but also the \so model we have followed on the ground of satisfying \astrong and proton decay constraints.

\subsection{The \astrong\,and proton decay constraints}
\label{astrongproton}


Starting with constraint related to the electroweak precision measurements, the effective weak - mixing angle $ \, \overline{sin}^2_f \hat{\theta} \; $ appears to be in a satisfactory level, with error less than $3 \sigma$, in comparison to its experimental value \cite{pdb}:
$
sin^2 \theta^f_{eff} \, = \, 0.23146 \, \pm \, 0.00012, 
$
over all the parameter space.
Thus, $ \, \overline{sin}^2_f \hat{\theta} \; $ does not form a substantial constraint for our analysis.

On the contrary, satisfaction of the experimental  bounds on the strong coupling constant \astrong, even by itself, imposes severe constraints and in conjunction with those from the proton lifetime reduces remarkably the soft parameter space.

Indeed, with the purpose of keeping \astrong within the experimental limits \cite{pdb}: 
$ \linebreak
\alpha_{strong} (M_Z) \, = \, 0.1184 \, \pm \, 0.0007,
$
with error $< \, 2 \sigma$, we should prevent $M_{1/2} \,$ and in particularly $\, m_0 \,$ from acquiring too large values.
Namely, we find that the \astrong~constraint is met successfully for $300 \, \lesssim \, m_0 \, \lesssim \,1400$ GeV and $500 \, \lesssim \, M_{1/2} \, \lesssim \, 1300$ GeV.
To achieve good results with the extreme high of the $\, m_0$ region we have to restrict $M_{1/2} \,$ between $800$ and $1100$ GeV at most. 
In respect for $M_{1/2} ,$ we can use its higher values, up to $1300$ GeV, if we choose small values for $\, m_0 ,$, around $500$ GeV. For further discussion 
refer to \cite{Katsikatsou:2010hj}.

This outcome is expected since for values of $\, m_0 \,$ and $M_{1/2} \,$ greater than $1$ ΤeV, the decoupling of supersymmetry is taken place. Thus, the Standard Model and MSSM end up with the same unsatisfactory predictions regarding grand unification of gauge couplings. 

\begin{figure}[t!]  
\begin{center}
\rotatebox{0}{\includegraphics[height=16cm,width=16cm]{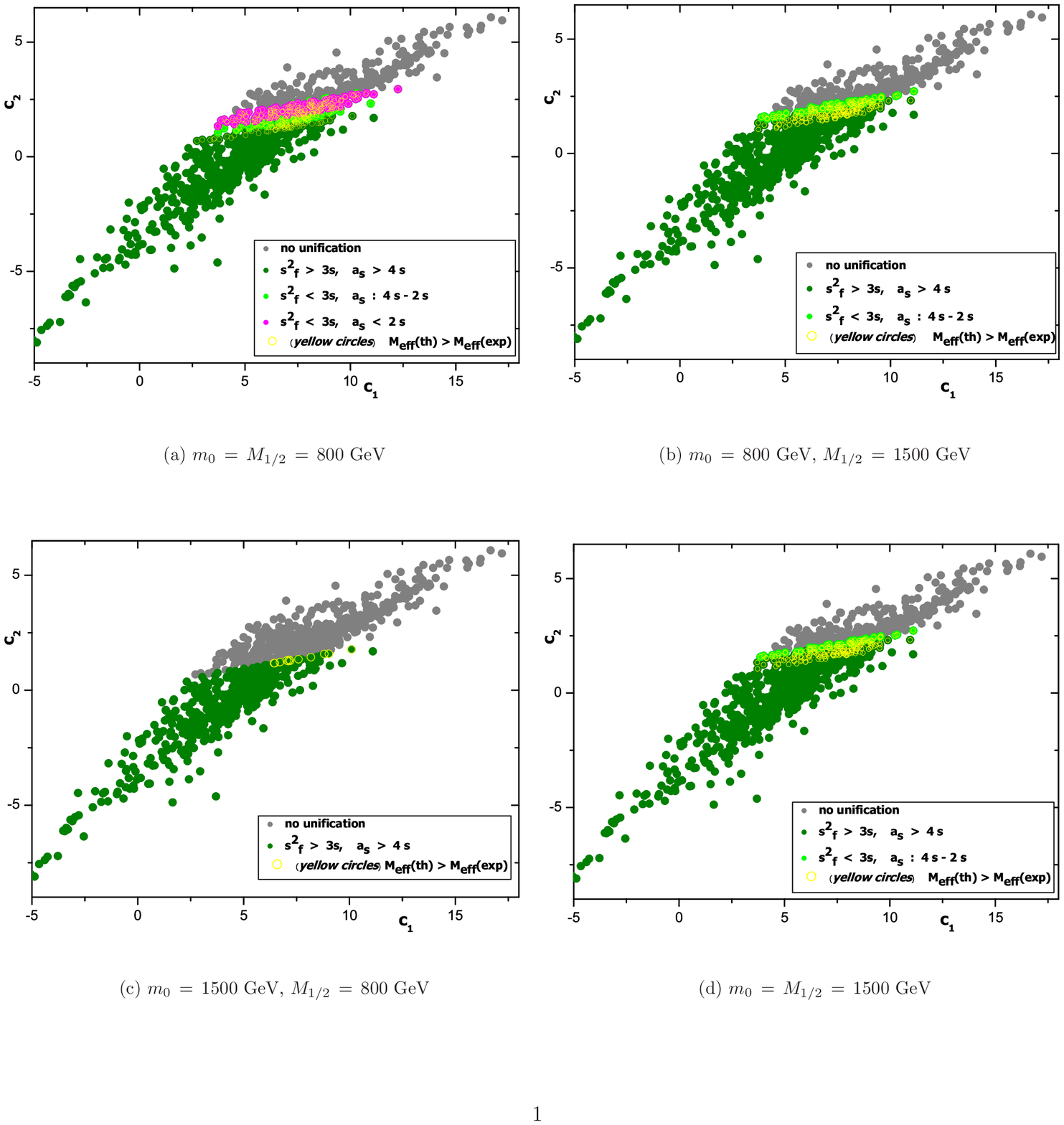}} 
\vspace{-1.5cm} 
\caption[]{\small  $\, M_{GUT} = 2 \cdot 10^{16} \,$ GeV, $\tan \beta \, = \, 10$ and $ A_0 \, = \, 100$ GeV, with $m_0, M_{1/2}$ as quoted. For a detailed description of the figure see main text.}
\label{fig_ap}  
 \end{center}
\end{figure}

Figure \ref{fig_ap} illustrates the pairs of $c_1, c_2$ as they are randomly generated and despite of that they are clearly correlated. Therefore, successful points ought to be within the diagonal stripe displayed in these figures.
The gray points represent those pairs of $c_1, c_2$ that fail even to give unification at the quoted \mgut, after the 2 - loop running of the  RGEs. For the green points, unification has been achieved but the value of \astrong is more than 
$4\sigma$ away. The light green points yield \astrong with error smaller than $4\sigma$, and the magenta region is the subset 
of magenta points corresponding to values of \astrong with the smallest possible error $< 2\sigma$. 
The yellow circles indicate the $c_1, c_2$ points which pass the proton decay constraint discussed below.

Regarding the proton decay constraint, we relay on the recent experimental lower bound of the proton lifetime (\ref{plifetime}) and we translate it into a lower bound of the mass parameter \meff, which controls the proton decay rate of eq.(\ref{width}), through the relation \ref{prtconstr}. Solving the RGEs, we perform this check for every randomly generated $\vec{c}\,$ and we find points that meet the constraint (\ref{prtconstr}). For these points, the predicted proton lifetime from our model lays over a region between $ \, 10^{34}$ and $ 10^{37} \,$ years.  We verify that the successful points overlap with the majority of the points that yield gauge coupling unification with values of \astrong within the $ 2\sigma\,$ experimental range. 

Raising of $m_0$ and/or $ M_{1/2}$ results to a reduction in the number of successful points around a central point. Besides, the unilateral increase of $m_0,$ while keeping $ M_{1/2}$ rather small, leads to experimentally unacceptable proton decay rates. We observe this for $m_0 \, = \, 900$ GeV and over, alongside with a difference from $ M_{1/2}$ at least equal to $700$ GeV
Furthermore, if we take small $ M_{1/2}$, $m_0$ should be no higher than $600 $ GeV from $ M_{1/2}$ to meet the proton decay constraint, e.g. $ M_{1/2} \, = \, 200 \,$GeV forces $m_0$ to be at most equal to $800 $ GeV. In addition, we observe that relation (\ref{prtconstr}) is spontaneously met when the 2 - loop RGEs yield \astrong with error smaller than $4\sigma$ for a given set of the parameters of the model.

%
\begin{figure}[h!]  
\begin{center}
\rotatebox{0}{\includegraphics[height=16cm,width=16cm]{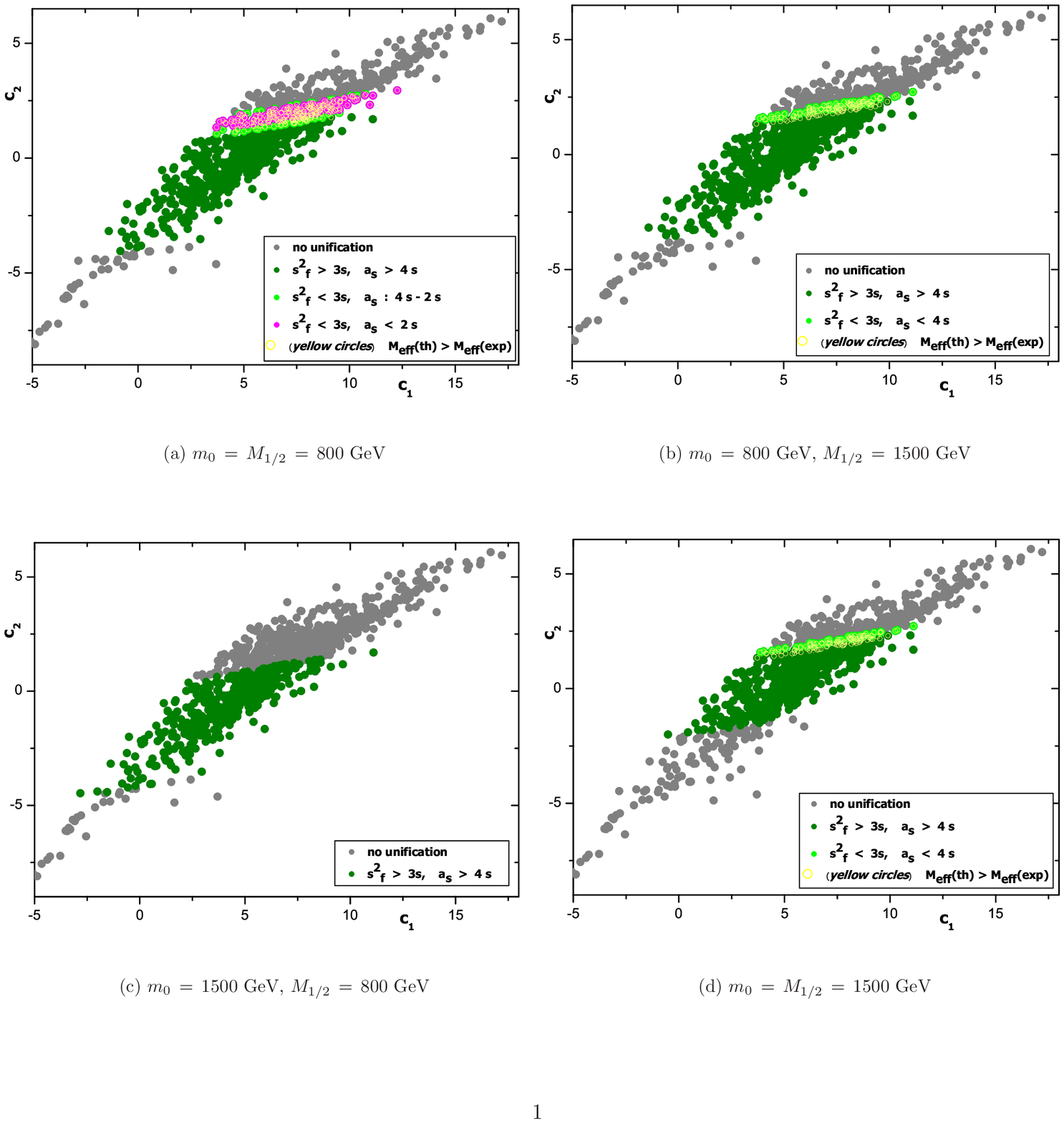}} 
\vspace{-1.5cm} 
\caption[]{\small  $\, M_{GUT} = 2 \cdot 10^{16} \,$ GeV, $\tan \beta \, = \, 45$ and $ A_0 \, = \, 100$ GeV, with $m_0, M_{1/2}$ as quoted. For a detailed description of the figure see main text.}
\label{fig_tb}  
 \end{center}
\end{figure}

\subsection{Dependence on \mgut, $\tan \beta$ and $A_0$}

As expected the value of the unification scale \mgut\,influences the results. In fact, by pushing \mgut\,to higher value provides easier satisfaction of both the \astrong and proton decay constraint. However, perturbativity limits on Yukawa and gauge couplings poses upper bounds on higher $\, M_{GUT}$ values \cite{Lahanas,Yanagida,Chang:2004pb} and hence we prefer the rather common value since LEP era \cite{Amaldi:1991cn,Ellis:1990wk,Langacker:1991an}:
 $\, M_{GUT} = 2 \,\cdot \, 10^{16} \, GeV$.
 
Another factor which affects our findings is the value of $\tan \beta$. A change of $\tan \beta$ from 10 to 45 causes a small decrease in the number of points which succeed to give unification. On the other hand, with a large $\tan \beta$ the number of points which give \astrong with error less than $4\sigma$ (or $2\sigma$ for $m_0 \, = \, M_{1/2}$ = $800$ GeV) slightly increases.
As far as the proton decay constraint is concerned, the points which satisfy (\ref{prtconstr}) show a considerable decrease, which starts from 22\% for $m_0 \, = \, M_{1/2} \, = \, 800$ GeV and reaches a 100\% for $m_0 \, = \, 1500$ GeV and $M_{1/2} \, = \, 800$ GeV. This was rather expected since $B_i$ in (\ref{prtconstr}) depends on $\frac{1}{\sin 2\beta}$. Hence, a change of $\tan \beta$ from 10 to 45 quintuples or so the values of \meffexp leaving, at the same time, the values of \meffth almost unchanged. These remarks for $\tan \beta \, = \, 45 $ are displayed in figure \ref{fig_tb}, in comparison with figure \ref{fig_ap} for $\tan \beta \, = \, 10. $ 

As far as $ A_0 $ is concerned, this parameter seems to play a moderate role in our analysis. If we stabilize the values of $m_0, \, M_{1/2}$ at their central region of values, as $m_0 =  M_{1/2} = 800$ GeV, 
we conclude that small positive values of $A_0$ marginally favour the fulfillment of proton lifetime restriction and also the emergence of \astrong values with less than $2 \sigma$ error. For $A_0 \, = \, - 3000$ GeV, a value inspired from models which enforce Yukawa unification \cite{pro_bdr}, we observe a small decrease in the success rate of those two constraints compared with the choice of $A_0 = \, 0$ and a behavior which approximates that of $A_0 = \, 1000$ GeV. 

\subsection{The $x$ parameter}

The random point samples that we use, define slices in the vector $\vec{c}\,$ space, for which the ratio $x$ (\ref{sh_param}) is constant. For the greater part of our analysis it is $x \, = \, 5$. This choice means a difference of one order of magnitude between the vevs of the superheavy Higgs fields in the adjoint $\vev{A}$ and the spinorial representation $\vev{C}.$ 
 
We perform the previously explained numerically procedure for different values of $x$, with an extended range, from $x \, = \, 0.5 \,$ up to $x \, = \, 500. \,$ Our conclusions are: First, for small $x$ the unification of gauge coupling is achieved by a larger number of randomly generated points $\vev{c}.$ Second, for large values of $x$, say $x \, = \, 500$, the proton decay constraint is satisfied ευκολότερα, while for very small values, for example $x \, = \, 0.5$, not even one point gives appropriate results to meet (\ref{prtconstr}). Finally, as far as \astrong is concerned, central values for $x$ favour the emergence of points which lead to \astrong with error less than $2 \sigma$. These remarks support the choice of $x = 5$ as the primary value in our analysis.

\subsection{Higgs and sparticles masses}

An important factor that nowadays every supersymmetric model has to take into account is the given results and limits set by the LHC experiment for the masses of the superparticles and also for the mass and the properties of the neutral Higgs boson. 

One of the biggest findings of LHC is the discovery of a particle with mass $\sim 126$ GeV, as ATLAS \cite{Aad:2012tfa} and CMS experiments \cite{Chatrchyan:2012ufa} has announced, in proton - proton (pp) collisions with centre-of-mass energy $\sqrt{s} \, = \,  8$ TeV. This particle has spin equal to zero and mainly positive parity couplings. These are characteristics that a Higgs boson is expected to have.

Gluinos are essential for the SUSY searches at LHC. Assuming $R$ parity conservation, sparticles are produced in pairs.
The strong production of first and second generation squarks and/or gluinos, that is to say the production of a pair of squarks or a pair of gluinos or the production of squark - gluino, is the supersymmetric process that is expected to dominate LHC searches. If squarks and gluinos are not too heavy, the strong production achieves its greatest cross section and hence permits the targeting of these (heavy) sparticles. Their subsequent decay is carried through cascades which are characterized by small branching ratios and usually long decay chains. These chains end up with the production of the LSP, which is considered to be stable, if $R$ parity is conserved.

The current exclusion limit of the gluino mass, which the ATLAS experiment lays \cite{TheATLAScollaboration:2013tha}, in the case of CMSSM, is $m_{\tilde{g}} \gtrsim 1300$ GeV, independently of the squarks mass. From the CMS experiment \cite{Chatrchyan:2014goa} the corresponding limit reaches the 1350 GeV, on condition that the gluinos and squarks has equal masses. For $\tan \beta \, = \, 30, \, A_0 \, = \, - 2 m_0$ and $\mu \, > \, 0$, all the gluinos with mass smaller than 1700 GeV are excluded, for values of $m_0$ up to 6 ΤeV, provided they have mass equal with the mass of squarks \cite{Aad:2014wea}. In the context of simplified models \cite{TheATLAScollaboration:2013tha,Chatrchyan:2014goa,Aad:2014wea,CMS:2013cfa}, it emerges that $m_{\tilde{g}} \gtrsim 1200 - 1300$ GeV, whereas the lower limit for the squarks mass can climb to about 900 GeV, depending on the simplified model in use. The tendency for the lower limit of gluino mass is upwards, judging from previous announcements from the ATLAS and CMS collaborations. Therefore we expect these quoted values to increase during the future runnings of the LHC. 

In the frameword of our analysis, with $\tan \beta \, = \, 10$ and $ A_0 \, = \, 100$ GeV, taken $m_0, \,=\, M_{1/2} \,=\, 800$ GeV and with the GUT scale fixed at  $ \mgut \, = \, 2\cdot10^{16}$ GeV, values which serve adequately both the \astrong constraint and the proton decay one, the running of the RGEs yields the lighter, neutral Higgs with mass $m_h \,=\, 125.55 \,\pm\, 0.05$ GeV (with $\sigma = 1.8$ GeV). This value is in agreement with the results for the mass of the scalar particle discovered at the LHC. Moreover the gluino mass emerges to be
$m_{\tilde{g}} \,=\, 1699.3 \,\pm\, 2.6$ GeV. The masses of first and second generation squarks occur at the same range. Positively, these derived values are way larger than the lower experimental limits placed by the LHC.

As expected, the increase in the value of $M_{1/2}$ affects almost proportionally the derived gluino mass. Thus, if we set $M_{1/2} = 1500$ GeV, leaving the rest of the soft parameters with the same values as in the previous paragraph, we get $m_{\tilde{g}} \,=\, 2972 \,\pm\, 6$ GeV. This result provides us with a safety net for a future potential raise in the exclusion limit of supersymmetry, after the next running of the LHC. However, this change in the value of $M_{1/2}$ causes a simultaneous increment in the lightest Higgs mass, along with that caused in the gluino mass. Hence, the output Higgs mass becomes almost $130$ GeV. On the contrary, by increasing $m_0$ the previously mentioned results are not affected. Finally, by setting $Α_0 = 1000$ GeV, while keeping the values of the other parameters mentioned above unchanged, we don't notice any important alterations in the results quoted in the previous paragraph, except for a small raise in the mass of the lightest Higgs, which now reaches $\sim 127$ GeV. If we seriously lower $Α_0$ to $- 3000$ GeV, the Higgs mass undergoes a reduction and becomes $\sim 122$ GeV. For the squarks and gluino mass there is no notable variation. We have to mention that in every case described, the LSP arises to be the lightest neutralino and in particular a pure bino.

There exists a particular set of values of soft parameters with $\tan \beta \, = \, 30, \, A_0 \, = \, - m_0$ and $\mu \, > \, 0$, considered in \cite{Aad:2014wea}, in the context of CMSSM. We investigate it using the values $m_0 \, = \, Μ_{1/2} \, = \, 800$ GeV which are proved to be the most favourable in respect of the satisfaction of constraints imposed by our analysis. With these values we end up with results which coincide the current exclusion limit  for gluino mass, since on average we get $m_{\tilde{g}} \,=\, 1750$ GeV. Moreover, the lightest Higgs mass emerges on average equal to 125 GeV. These finding urge us to a more detailed examination of the soft parameters space.

\section{Conclusions} \label{conclusions}

Our goal is to check the viability of SUSY GUTs, using electroweak precision and proton decay data, as well as recent results from the LHC experiment.
We present a new method according to which the effects of the HET, in a GUT model, can be described collectively by fewer, carefully selected parameters that are randomly produced from the original set of the numerous parameters of the model. In this way, the scanning over the parameter space for favourable regions, in accord with the experimental data, becomes easier and less time-consuming. Moreover, this method, as developed, can be applied to any GUT model, regardless of its complexity.

To check the efficiency of this method, we directly apply it to a SUSY SO(10) GUT model, in which the doublet-triplet splitting is realized through the Dimopoulos-Wilczek mechanism. Only five parameters ($c_{1, 2, 3}, \,  \tan \theta$ και$ M_L$), randomly generated from the superheavy spectrum, incorporate the large number of HETs which this model provides, in the context of CMSSM.  

We show that there exists regions in the space of these parameters which satisfy all the addressed constraints. These regions are endorsed by small to central values of $m_0$ and $M_{1/2}$ from $500$ GeV up to $1.5$ TeV. The value of $\tan \beta$ affects mainly the proton decay constraint and must be kept small to moderate. We also note that small positive values of $A_0$ enforce the success of our analysis.
Finally we take into consideration the recent results from the LHC experiment concerning the Higgs mass and the exclusion limits of supersymmetry. We demonstrate that the output supersymmetric masses by our model favour the choice of central values in the $m_0 - Μ_{1/2}$ plain. However a further investigation is needed regarding different regions in the soft parameters space.

In this way, we figure out that the method is both convenient and efficient and the SUSY \so theory remains a prominent extension of the Standard Model.

\section*{Acknowledgements}
The author is grateful to A.B. Lahanas for extensive discussions, critically reading 
the manuscript and continuous support during this effort.
The author wishes to acknowledge partial support from  the University of Athens Special Research Account.



\end{document}